# Resonances of the magneto-optical intensity effect mediated by interaction of different modes in a hybrid magnetoplasmonic heterostructure with gold nanoparticles

**Anastasiya E. Khramova[1,2], Daria O. Ignatyeva[1,2], Mikhail A. Kozhaev[1,3], Sarkis A. Dagesyan[2], Vladimir N. Berzhansky[4], Aleksander N. Shaposhnikov[4], Sergey V. Tomilin[4], Vladimir I. Belotelov[1,2]**

[1] *Russian Quantum Center, 45, Skolkovskoye shosse, Moscow, 121353, Russia*
[2] *Faculty of Physics, Lomonosov Moscow State University, Leninskie Gory, Moscow 119991, Russia*
[3] *Prokhorov General Physics Institute of the Russian Academy of Scien, 38 Vavilov Street, Moscow 119991, Russia*
[4] *Vernadsky Crimean Federal University, 4 Vernadskogo Prospekt, Simferopol, 295007, Russia*
*\*ae.khramova@gmail.com*

**Abstract:** Here we demonstrate a novel magnetoplasmonic heterostructure for efficient control of light. It consists of gold nanoparticles embedded in a thin magnetic film covered with a gold layer pierced with periodic nanoslit array. Unique feature of the proposed structure is that it supports four different types of optical modes in the same frequency range including localized and propagating surface plasmons along with waveguide modes. A peculiar magneto-optical response appears at the frequencies of the mode hybridization. The most important result comes from hybridization of the localized and propagating plasmons leading to a significant increase of the magneto-optical effect intensity.

## 1. Introduction

Recently, magneto-optical methods for control of light have been actively implemented, and one of the promising approaches is the utilization of the magnetoplasmonic structures [1-9]. The magnetoplasmonic structures promising for ultracompact optical isolators [10,11], highly sensitive magnetic sensors [12] and biosensors [13-15]. However, enhancement of the magneto-optical response of the plasmonic structures remains an important issue since it might expand their application range. Further progress in this direction can be made through interaction of different types of modes excited in the structures simultaneously [14-16].

Propagating surface plasmon-polaritons (SPP) are sensitive to a magnetic field applied in the transverse direction, which is accompanied by pronounced resonances in the magneto-optical intensity effects, the transverse Kerr effect (TMOKE) [17]. On the contrary, the localized plasmonic resonances (LSP) are known to enhance the magneto-optical polarization effects like Faraday effect or longitudinal magneto-optical Kerr effect [18,19], while are marginally influenced by the transverse magnetic field and provide only weak peculiarities in the TMOKE spectra .

It was previously shown that by combining in one structure an array of gold nanoparticles with a dielectric film, one can get a strong interaction between LSP and SPP, as is indicated by notable anticrossing of resonances in the reflectance spectra [20]. In an external magnetic field such coupling of LSP and SPP modes provides spectral shift of both resonances thus making LSP also sensitive to the transverse magnetic field and mediating the TMOKE increase.

Another interesting type of mode interaction where LSP plays a notable role was demonstrated in the structure consisting of bismuth iron-garnet thin film and gold nanowires [21]. Properly designed structure allowed enhancement of the TMOKE due to the hybridization of LSP with a waveguide mode (WG) in the iron-garnet film.

In this work we consider modification and enhancement of the intensity magneto-optical effects in a novel type of the hybrid magnetoplasmonic structure formed by a thin bismuth-substituted iron-garnet layer with nanoparticles embedded inside it and gold gratings deposited on its surface. Therefore, several types of modes could be excited in this structure: LSPs in the nanoparticles, WG modes in the iron-garnet layer, SPPs at the interfaces of the gold grating and the iron-garnet or air. Due to their different dispersion, some of these modes could be excited simultaneously providing their hybridization and significant modification of the magneto-optical response observed for the transversal magnetization of the iron-garnet layer. Since this magnetooptical response is observed in transmission the corresponding variation of the transmittance $T(M)$ caused by the magnetization reversal can be referred as TMOKE in transmission defined by: $\delta = (T(+M)-T(-M))/(T(+M)+T(-M))$. This magneto-optical intensity effect in transmission has the same nature and is determined analogously as the conventionally observed TMOKE in reflection.

In contrast to previously reported structures [16,18], the proposed magnetoplasmonic structure supports four different types of modes at different wavelengths and angles, that could be controlled independently by variation of structure parameters. It will be shown that their hybridization could produce both significant enhancement or noticeable suppression of TMOKE.

## 2. Modes in hybrid magnetoplasmonic structure with gold nanoparticles

In order to observe the effects of coupling and hybridization of different optical modes the following magnetoplasmonic structure was fabricated, Fig. 1(a). An ultrathin gold wedge with thickness increasing from 1 nm to 15 nm was deposited by thermal evaporation on a gallium-gadolinium garnet (GGG) substrate at 150°C using original "thin screen" method [22]. For the formation of the nanostructures the samples were annealed in air at atmospheric pressure at temperature 950ºC during 10 min. It resulted in the formation of nanoparticles (NPs) with characteristic size variation along the sample [23]. Table 1 presents data for three points along the wedge thickness gradient C1, C2, C3. (Fig. 1). Average size of the gold nanoparticles in these points varies from 55 nm at C1 to 78 nm at C3. After formation of the gold nanoparticles, the sample was coated with 100-nm-thick bismuth-substituted iron-garnet (BIG) using reactive ion-beam sputtering technique in argon–oxygen mixture and annealed at temperature of 680 °C during 20 min. More information on fabrication process and sample characterization is given in [23]. The iron-garnets with Bi substitution have relatively large magneto-optical parameter that makes them one of the main materials for magnetoplasmonics [24,25]. For excitation of the propagating SPPs and WG modes 80-nm-thick gold gratings were deposited on top of the BIG film at C1, C2, and C3 points along the sample. At each point two gratings were fabricated with periods of $P$=550 nm, $P$=510 nm and the air slit width of 90 nm.

**Table 1.**

| Point along the wedge | The average size of nanoparticles (nm) | Nanoparticle standard deviation | LSP resonance $\lambda_1$ (nm) | LSP resonance width (nm) |
|---|---|---|---|---|
| 1 | 55 | 18 | 680 | 150 |
| 2 | 65 | 22 | 700 | 177 |

| 3 | 78 | 27 | 740 | 222 |

Firstly, we consider modes which could be excited in the hybrid magnetoplasmonic heterostructure and the features of their magnetooptical response (Fig. 1(b)).

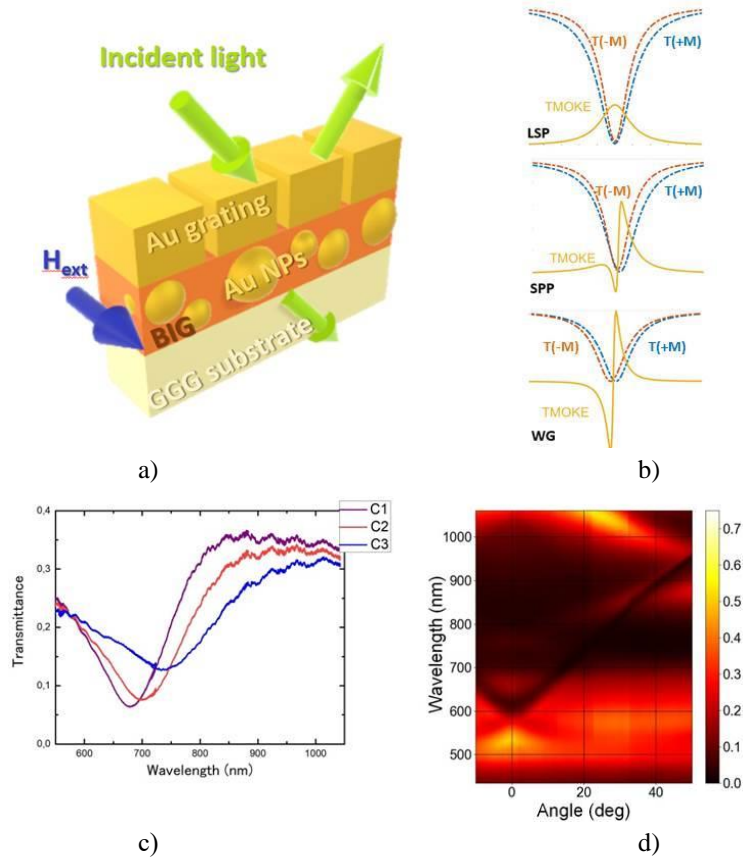

Fig. 1. Magnetoplasmonic heterostructure with Au nanoparticles covered with a 1D gold grating. (a) Sample design. (b) Schematic representation of TMOKE shape for different types of modes of a nanostructured sample based on thin magnetic dielectric film. The red and blue lines indicate the transmittance spectra of the sample with the forward and backward directions of the external magnetic field, orange line shows the shape of the TMOKE response. (c) Transmission spectra for uncoated BIG film with nanoparticles of different size: $D$=55 nm, $D$=65 nm, $D$=78 nm (d) Typical transmission spectra of the hybrid magnetoplasmonic structure with Au nanoparticles and grating (D=65 nm, P=550 nm).

1.  *Localized surface plasmon (LSP) in gold nanoparticles*. The LSP is observed by a broad deep in the transmittance spectra (Fig. 1(c)). The transmittance spectra in Fig. 1(c) were measured for light passing through the parts of the sample without gold gratings. At the considered points on the sample the LSP resonance shifts from $\lambda_{c1}$ =680 to $\lambda_{c3}$ =740nm (Table I). The spectral position of the resonance is known to be determined by the shape, size and materials and is independent on the angle of incidence. The experimentally measured wavelength of the LSP is slightly shifted towards lower wavelengths compared to the value determined by equation $\varepsilon_{Au}/\varepsilon_{BIG} = -2 - \frac{12}{5}(k_0 n_{BIG} a)^2 - 2i(k_0 n_{BIG} a)^3$ [26]. It is explained by LSP penetration inside the GGG substrate and air making the effective refractive index of the surrounding media smaller: $n_{BIG}^{eff} = 2.32$ instead of $n_{BIG} = 2.38$ at wavelength of 750 nm.

Apart from the resonance shift the width of the resonances also varies from 150 nm for $\lambda_{c1}$ to almost 250 nm for $\lambda_{c3}$. The broadening is mostly due to the method of the nanoparticle formation which provides rather wide range of nanoparticle sizes. This leads to the additional broadening of the LSP resonances, especially for larger nanoparticles. Different resonances for nanoparticles of different size allow us to observe hybridization and interaction of different modes with LSP in a wide spectral range.

The TMOKE is usually marginally influenced by LSP. It is explained by nanoparticle polarizability variation [19] caused by the external magnetic field. The LSP resonance position does not experience shift, at the same time its magnitude and width are modified, making TMOKE for LSP modes to be bell-shaped rather than S-shaped (Fig. 1(b)).

2. *Propagating surface plasmon polaritons (SPPs) in the gold grating at air-gold and BIG-gold interfaces* (called further air-SPP and BIG-SPP, respectively). SPPs are known to be excited at a metal-dielectric interface with a wavenumber $\beta = k_0 \sqrt{\frac{\varepsilon_m \varepsilon_d}{\varepsilon_m + \varepsilon_d}}$ exceeding that of the incident wave. This requires usage of coupling elements, such as diffraction gratins, which could excite SPPs under the fulfillment of phase matching condition: $k_0 \sin\theta + \frac{2\pi}{P} m = \beta$ (in the empty-lattice approximation), where $P$ is the grating period, $\theta$ is the angle of light incidence and $m$ is an integer corresponding to the evanescent diffraction order exciting the surface wave. Both parallel and anti-parallel directions of the SPP wavevector and tangential component of the incident light could be achieved via coupling of incident light to the diffraction orders of opposite signs forming in the light-wavelength – angle-of-incidence coordinates two branches in a form of the cross symmetrical with respect to $\theta = 0°$ line (Fig. 1(d)).

A transversal magnetic field significantly modifies both real and imaginary parts of the propagation constant of the BIG-SPP making the SPP-associated resonances spectrally shifted due to the variation of Re($\beta$) and change form due to the variation of Im($\beta$) with the magnetization reversal (Fig. 1(b)). Therefore, TMOKE resonance has S-shape with positive (negative) peak much larger (lower) than the opposite one, and this ratio reverses for the opposite SPP propagation directions. Optical field of the excited air-SPP is concentrated near air-gold interface thus diminishing the energy inside the BIG layer. This makes observed TMOKE for this mode significantly weaker than for the BIG-SPP.

3. *Waveguide (WG) modes in the BIG layer*. A WG mode could be excited inside the BIG layer, since it experiences total internal reflection from the lower index GGG substrate and is reflected from the Au grating serving as the coupling element for the WG mode excitation as well. Dispersion of the WG mode is also nonreciprocally affected by the transversal external magnetic field, however the magnetically-induced variations of $\beta$ are smaller than for the BIG-SPP mode. This results from the WG mode field distribution concentrated in the central part of the film which makes these modes less sensitive to such surface magneto-optical effect as TMOKE than the BIG-SPPs whose energy is concentrated near the BIG surface.

In planar all-dielectric structures attenuation of the WG mode is small, and the TMOKE resonance is determined mainly by the spectral shift due to modification of Re($\beta$) at the WG mode which leads to pure S-shape of the TMOKE resonance [17] (Fig. 1(b)). At the same time, WG modes excited in the metal-dielectric structures have significantly broader resonances due to higher Im($\beta$) and therefore they provide TMOKE peculiarities mainly by the modification of the shape of the resonance and the TMOKE resonance acquires non-symmetric S-shape [27]. In our case, the damping of the WG mode is increased even more due to the presence of gold nanoparticles inside of the iron-garnet layer.

The fabricated hybrid magnetoplasmonic structures with Au nanoparticles and gratings allow one to excite all four types of modes listed above. Moreover, since these modes are characterized by different dispersion, one could trace their interaction and reveal the impact of the mode hybridization on the magneto-optical response.

## 3. Mode interaction and TMOKE enhancement in the hybrid structures

Now we trace the mode interaction and its impact on the magneto-optical response of the hybrid structure. Typical spectra of the TMOKE response is shown in Figs. 2(a), 2(c) and 2(e), while the plots in Figs. 2(b), 2(d) and 2(f) illustrate the dispersion of the modes calculated in the empty-lattice approximation for the propagating SPP and WG modes, and estimated from the experimental results (Table I) for the LSP modes.

The most prominent and wide TMOKE cross in these graphs refers to the BIG-SPP mode excited via $m=\pm2$ diffraction order of the grating (red lines with cross center at 790 nm in in Figs. 2(b), 2(d) and 2(f) and at 750 nm in Fig. 2(h)). Notice that TMOKE vanishes in two regions of this color plot: at $\theta=0°$ and at around $\theta=41°$ (refer to orange region in Figs. 2(b), 2(d), 2(f) and 2(h)). TMOKE always vanishes at normal incidence due to symmetry reasons. However, zero TMOKE at high incidence angles (orange region in Figs. 2(b), 2(d), 2(f) and 2(h)) has different origin, since it is related to the edge of the first Brillouin zone: $k_0 \sin\theta = \pi/P$ (black dashed lines in Figs. 2(b), 2(d), 2(f) and 2(h)).

The dispersion curves of SPPs propagating in opposite directions that excited by different diffraction orders $m = \pm2$ and $m = \pm3$ intersect, which means their simultaneous excitation. Their magneto-optical responses have different sign, as it was mentioned before, and nearly compensate each other. Small background arises from the different efficiency of excitation of modes via different diffraction orders.

Also, one could see how the dispersion of the BIG-SPP excited by diffraction order $m = \pm3$ (see Figs. 2(b), 2(d), 2(f) and 2(h)) becomes almost horizontal asymptotically reaching plasmonic wavelength $\lambda_{SPP} \sim 700\ nm$ (where $\beta_{SPP} \to \infty$ for this structure). Notice also that this resonance is much weaker than the BIG-SPP resonance corresponding to $m = 2$.

The LSP (which position is shown by green line in Figs. 2(b), 2(d), 2(f) and 2(h)), as it was discussed above, exhibits very small bell-shaped TMOKE resonance. Pure LSP without interaction with other modes is poorly seen in the TMOKE spectra (Fig. 2(c), LSP position is shown by green line in Fig. 2(d)). However, if Au NPs embedded inside the garnet layer are covered with a metal grating, the strong interaction between BIG-SPP and LSP occurs also affecting the structure response to the transversal magnetization, see below. The mode hybridization leads to two different variations of the TMOKE spectra.

First, if the signs of the TMOKE for LSP and BIG-SPP coincide ($m = \pm3$ in our structure), one may observe the significant increase of the TMOKE caused by the mode hybridization (Fig. 2(a), refer to purple region in Fig. 2(b)). Such amplification is explained by two reasons. On the one hand, the BIG-SPP has propagating character allowing for the non-reciprocal wavenumber variation; on the other hand, the penetration inside the magnetic layer increases due to hybridization with the LSP mode. For the nanoparticles providing LSP at wavelengths higher (green line in Fig. 2(f)) or lower (green line in Fig. 2(d)) than the BIG-SPP wavelength (red line with $m=3$ in corresponding figures) the TMOKE response of both the LSP and the BIG-SPP is almost one order lower (Figs. 2(c) and 2(e)). It is also worth noting that the BIG-SPP resonance is deteriorating due to the increase of the NP diameter increase followed by increase of the scattering inside the magnetic layer (see magnitude of the BIG-SPPs with $m=2$ in Fig. 2(e) compared to Figs. 2(a), 2(c)).

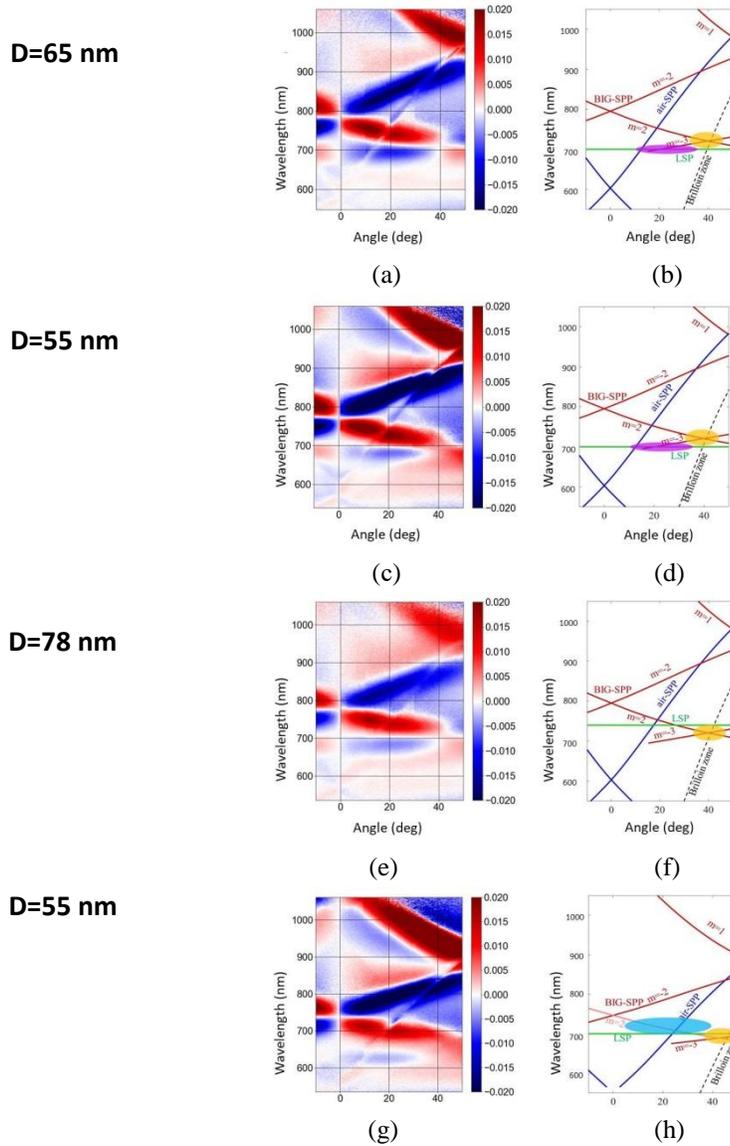

Fig. 2. (a),(c),(e),(g) Experimental wavelength vs. angle spectra of TMOKE for the hybrid magnetoplasmonic structure with Au NPs and 1D gratings and (b),(d),(f),(h) mode dispersion for the corresponding sample configurations. BIG-SPP, air-SPP and LSP modes are shown with red, blue and green curves, correspondingly. Black dashed lines show the edge of the first Brillouin zone. Orange region show disappearance of TMOKE near the edge of the Brilloin zone. Purple region shows an increase in TMOKE caused by mode hybridization. Blue region show the attenuation of TMOKE via LSP and BIG-SPP interaction in the case of different TMOKE sign of each of the branches. Experimental and numerical spectra are obtained for the samples with the following NPs size and grating periods: (a-b) D=65 nm, P=550 nm, (c-d) D=55 nm, P=550 nm. (e-f) D=78 nm, P=550 nm.(g-h) D=55 nm, P=510 nm.

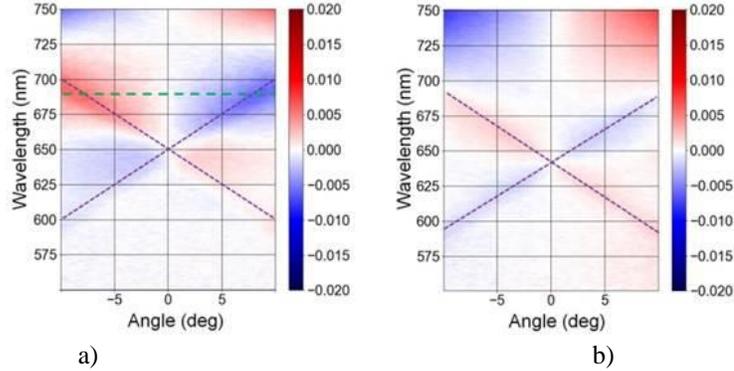

a)                                                                b)

Fig. 3. Experimental wavelength vs. angle spectra of TMOKE enhanced via excitation of WG mode in the hybrid magnetoplasmonic structure with Au nanoparticles and 1D grating. (a) Enhancement of TMOKE via interaction of the WG mode (purple dashed line) and LSP (green line). (b) TMOKE for a pure WG mode (purple dashed line).

     Second possibility is the interaction of the LSP and BIG-SPP branches of different TMOKE sign (light blue region in Fig. 2(h)). In this case, the interaction of the LSP with the BIG-SPP is noticeable and results in a significant suppression of the TMOKE (Fig. 2(g)). The lower branch of the BIG-SPP (corresponding to $m=+2$, light red line in Fig. 2(h)) meeting the LSP (green line in Fig. 2(h)) becomes much weaker than the upper one ($m=-2$, dark red line in Fig. 2(h)). As the LSP itself has a very small TMOKE more than one order lower than TMOKE of BIG-SPP for $m=\pm 2$ mode, as it was shown above, we underline that this TMOKE suppression is the result of the mode hybridization rather than being the mathematical sum of the two opposite-sign effects.

     At the same time, peculiar magnetooptical interaction pattern is observed in the air-SPP (shown by dark blue lines in Figs. 2(b), 2(d), 2(f) and 2(h)) and the BIG-SPP interaction. Mode symmetry requires very strong anti-crossing of the BIG-SPP and the air-SPP, which is most clearly seen at -40 deg, 900 nm (Fig. 2(a)).

     Another interesting possibility is the hybridization of the LSP and WG mode, see Fig. 3(a). The presence of NPs inside the waveguide BIG film significantly increases scattering losses of the WG mode and its absorption so it is rather weak both in transmittance and TMOKE spectra. Hybridization of the WG mode (shown by purple dotted cross in Fig. 3(a)) with the LSP (green dotted line in Fig. 3(a)) results in the increase of TMOKE. This enhancement makes the TMOKE for hybrid LSP-WG branch (top of the purple cross, Fig. 3(a)) noticeably more intense than for the pure WG branch (bottom part of the purple cross, Fig. 3(a)). Similar structure with the shifted to higher wavelengths LSP resonance demonstrates equal TMOKE magnitude for both WG mode branches (LSP wavelength is 740 nm for structure in Fig. 3(b)).

     Full transmittance spectra for lattice represent in appendix [28].

## 4. Conclusion

We performed a comprehensive study of the TMOKE in the hybrid magnetoplasmonic heterostructures consisting of gold nanoparticles embedded in a thin magnetic layer covered with gold gratings. Such structure is unique to support various types of modes in visible and near-IR range: localized plasmons in nanoparticles, propagating surface plasmon-polaritons at the air-gold and garnet-gold interfaces, and waveguide modes in the iron-garnet layer. All of these modes were observed experimentally both in transmittance and TMOKE spectra. We have demonstrated that TMOKE spectra allows observing peculiar pattern of mode interaction in a clearer way than transmittance spectra do. We have shown the possibility to enhance the weak TMOKE at the LSP resonance via the hybridization of the LSP with propagating SPP modes. The most prominent feature of this structure is a significant enhancement of the TMOKE for the hybrid LSP/BIG-SPP mode near the plasmonic frequency. It should be noted that TMOKE is almost zero for each of these two modes excited independently. Therefore, inclusion of the nanoparticles in the ferromagnetic layer could be used to control and enhance the magnetooptical response of the magnetoplasmonic structures.

**Funding**

We acknowledge the financial support of Russian Science Foundation (RSF) (17-72-20260) and the Basis Foundation (18-2-6-202-1) in the part of theoretical and experimental studies of the magnetoplasmonic structures. ANS and SVT acknowledge support by the RF Ministry of Education and Science in the framework of the State task (project no. 3.7126.2017/8.9) in the part of the sample fabrication.

**APPENDIX**

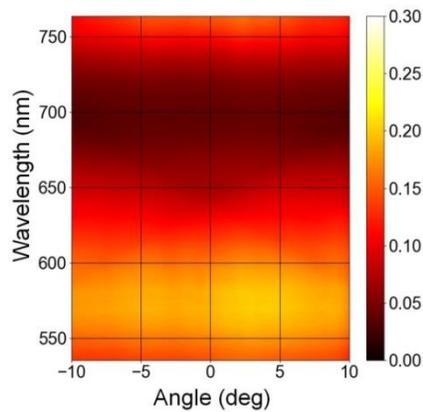

(a)

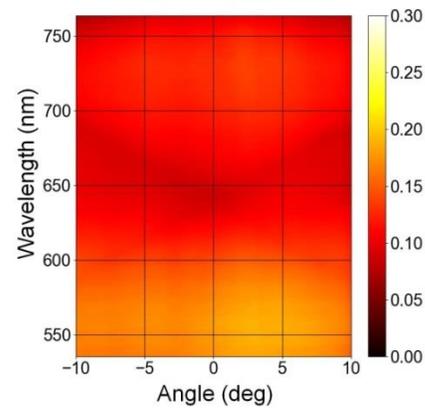

(b)

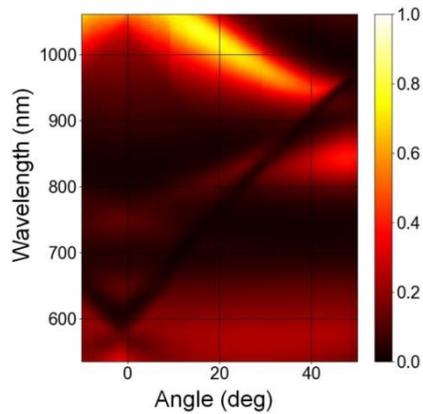

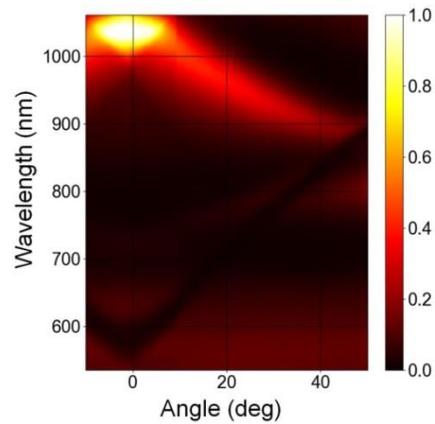

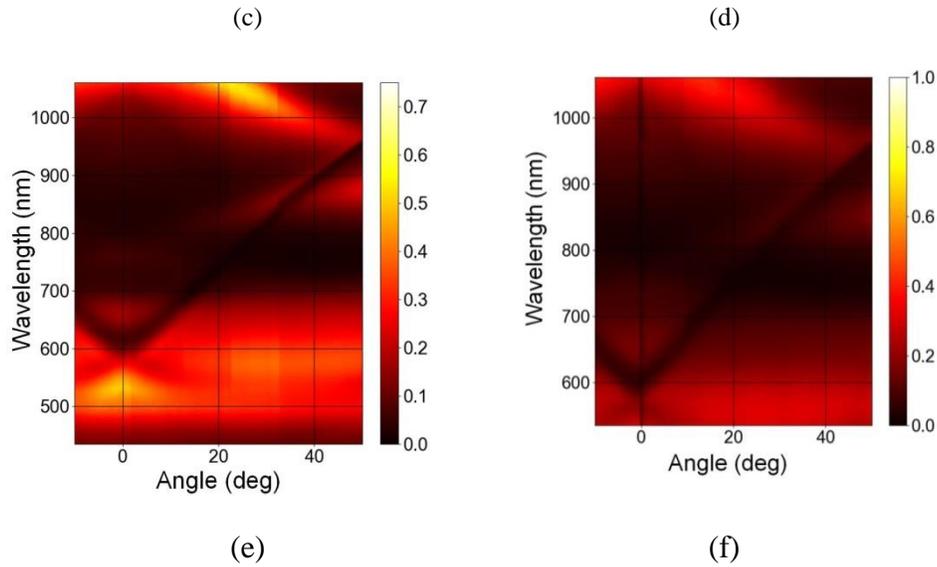

(e)                        (f)

Fig. 4. Experimental wavelength vs. angle transmittance spectra for hybrid magnetoplasmonic structure with Au NPs and 1D grating: (a) D=55 nm, P=330 nm. (b) D=78 nm, P=330 nm. (c) D=55 nm, P=550 nm. (d) D=55 nm, P=510 nm. (e) D=65 nm, P=550 nm. (f) =78 nm, P=550 nm. The modes seen in the spectra are defined in Fig. 2,3.

28. Appendix